\documentclass[aps,prl,preprint,showpacs]{revtex4}

\usepackage{graphicx}

\usepackage{color}

\newcommand{\be}{\begin{equation}}

\newcommand{\ee}{\end{equation}}

\newcommand{\ba}{\begin{eqnarray}}

\newcommand{\ea}{\end{eqnarray}}

\begin{document}


\title{Broad-band electron spectroscopy:
a novel concept based on Thomson scattering }

\author{ P. \, Tomassini}

\author{M.\, Galimberti}
\altaffiliation[Also at~]{Dipartimento di Fisica Universit\`a di Pisa, Unit\`a INFM, Pisa Italy}

\author{A.\, Giulietti}
\author{D.\, Giulietti}
\altaffiliation[Also at~]{Dipartimento di Fisica Universit\`a di Pisa, Unit\`a INFM, Pisa Italy}

\author{L.A. \, Gizzi}
\author{L. \, Labate}
\altaffiliation[Also at~]{Dipartimento di Fisica Universit\`a di Bologna, Bologna Italy}
\affiliation{Intense Laser Irradiation Laboratory - IPCF, Area della Ricerca CNR, Via Moruzzi 1, 56124 Pisa, Italy}

\pacs{13.60.Fz, 29.30.Dn, 41.75.Ht, 41.75.Jv}
\begin{abstract}
The  spectrum of relativistic electron bunches with large energy dispersion is hardly obtainable with conventional magnetic spectrometers.
  We present a novel spectroscopic concept, based on the analysis of the photons generated by Thomson Scattering  of a probe laser pulse inpinging with arbitrary incidence angle onto the electron bunch.
 The feasibility of a single-pulse spectrometer, using an energy-calibrated CCD device as detector, is investigated.  Numerical simulations performed in conditions typical of a real experiment show the effectiveness and accuracy of the new method.
\end{abstract}
\maketitle

Thomson Scattering of intense laser beams from  charged particles
 has attracted a great interest for a
variety of research areas, including
 particle acceleration \cite{Leemans1}, laser-plasma interactions \cite{Leemans2}
\cite{Renard} \cite{LaserPlasma}, plasma fusion \cite{fus1} \cite{fus2},
 X-rays generation \cite{Leemans} and medical diagnostics \cite{Medical}. In this letter we investigate the possibilty of using linear Thomson scattering of a laser beam inpinging at arbitrary incidence angle onto a relativistic electron bunch to design  a novel  electron energy spectrometer.  The basic idea is to measure the spectral and angular distribution  of the scattered laser pulse and use it to infer the energy spectrum of the electrons.  The concept can be in principle applied to a wide class of electron bunches, but it seems particularly attractive for the study of laser accelerated electrons.

Electron spectroscopy is currently performed with magnetic spectrometers \cite{Sp1}, whose use is practically limited to rather well collimated electron beams, with a moderate energy spread around a roughly known mean value.  The magnetic electron spectroscopy is much more difficult to apply to electron bunches having a broad band energy spectrum and  whose mean energy and degree of collimation are not well predictable.  Such conditions are typical, for example, of electron bunches produced by laser acceleration of plasma electrons \cite{Nostro} \cite{LWFA}. In this latter case, when magnetic spectrometers are used, the electron spectrum is usually  obtained with several laser shots, each one devoted to produce a portion of the whole spectrum, which results in a poor accuracy.  As the research on laser acceleration of electrons is rapidly growing, there is a need of a new class of spectrometers.

The computation of the Thomson scattered photon yield
 by a bunch of  charged particles is based on a classical electrodynamic approach
 \cite{Jackson}
and, in the case of {\it incoherent} scattering, it can be
expressed as an integration of the single electrons spectra
\cite{Leemans}:
\be
\label{Yield1}
{d^2 N\over d\omega d O }(\omega,\theta)
=   \int d{\bf x}d{\bf p} \, f({\bf x},{\bf p},t) {d^2 N\over d\omega d O }|_{Single}
\ee
\noindent where the general form of the single electron scattered photon yield ${d^2 N/d\omega d O }|_{Single}$ can be found in \cite{Esarey} and \cite{Ride}, $\omega$ is the pulsation of the scattered radiation,
$\theta$ is the scattering angle in a spherical coordinate system,
 ${\bf n}(\theta,\phi)$ is
the scattering direction, $dO\equiv d(\cos \theta) d\phi$ is the solid angle and $f({\bf x},{\bf p},t)$
 is the phase-space statistical distribution of the bunch.

Let us consider a relativistic electron bunch  moving in the
'Laboratory' frame along the $z$ direction, whose  statistical
distribution function is parametrized
 as
$f({\bf x},{\bf p},t)\equiv F(\gamma) \Theta_\gamma(\theta) n_\gamma({\bf r}-\beta c t)$,
beying $F(\gamma)$ the energy distribution function and  $\Theta_\gamma$, $ n_\gamma$
 describe the angular and spatial statistics of the population of electrons having  relativistic factor $\gamma$.

 A laser pulse of Gaussian longitudinal and transverse envelopes of duration $T$ and waist $w$, intensity $I_0$ and reduced vector potential of nonrelativistic amplitude  $a_0 = 8.5\cdot 10^{-10} \lambda [\mu m] \sqrt{I_0}[W/cm^2]<<1$,
propagates in the $x-z$ plane with angle $\alpha^L$ with respect
to $z$ and  intersects the electrons trajectories, thus  making
them oscillating  and irradiating.

In order to obtain a close expression of the photon yield
generated by Thomson Scattering of the laser pulse by the electron
bunch, we will assume very general  assumptions. First, the number
of cycles $N_0$ of the scattered  light emission by a single
electron is very large, so that the spectrum of the radiation
scattered at a certain angle by the  electron is monocromatic.
Second, the angular distribution of the bunch with collimation
angle $\Delta\theta_e $ is known (a collimator is introduced or
the bunch divergence is monitored e.g. with the method reported in
Ref. \cite{Leemans1}). Finally, each population of electrons
having the same energy has a Gaussian density distribution with
$\sigma_L, \, \sigma_T$ the  (energy dependent) longitudinal and
transversal bunch sizes, respectively, and position of the center
of the spatial distribution at $t=0$ denoted with $\zeta$.

 The evaluation of Eq. \ref{Yield1} with the assumptions reported above, yields the following expression for the emittance
\begin{widetext}
\ba
\label{Yield4}
{d^2 N\over d\omega d \cos\theta}(\omega,\theta)
&=&\alpha{\sqrt{2\pi} \over 8} a^2_0 \int d\gamma d\theta_e \Theta_\gamma(\theta_e)N_0(\gamma)F(\gamma)
\left({\cal H}(\gamma) e^{-{\cal K}(\gamma)(\Delta T-\zeta/\beta(\gamma))^2}\right)
\nonumber\\
&\times& {1\over \gamma^2 (1-\beta(\gamma) \cos(\theta-\theta_e)^2}
\delta\left(\omega - {\tilde \omega}(\theta-\theta_e,\gamma)\right)\, ,
\ea
\end{widetext}
\noindent where the 'resonant' scattered radiation pulsation
${\tilde \omega}(\theta,\gamma)$ is given by
$${\tilde \omega}(\theta,\gamma)=
 \Omega {(1-\beta(\gamma)\cos \alpha^L)\over
 (1-\beta(\gamma)\cos \theta)}\, ,$$
\noindent being $\Omega$ the pulsation of laser pulse. The
envelope functions ${\cal H},\, {\cal K}$ take into account the
spatial integration of the single electron photon yield and are
computed as \ba \label{HK} {\cal H}^{-2}&=&
\left(1+2{\sigma^2_T\over w^2}\right) \left(1+2{{\tilde T}^2\over
T^2} {\sigma^2_L\gamma^2\cos^2\delta\over w^2}\right)\times
\nonumber\\ &\times & \left(1+2{{\tilde T}^2\over T^2}{
\sigma^2_T\sin^2\delta\over w^2+2\gamma^2\sigma^2_L{{\tilde
T}^2\over T^2}\cos^2\delta}\right)\nonumber\\ {\cal K}&=&
2{{\tilde T}^2\over T^2} { (\gamma \beta c \cos\delta)^2\over w^2+ 2({\tilde T}^2/
T^2)\left( \gamma^2 \sigma^2_L \cos^2\delta+\sigma^2_T
\sin^2\delta\right)}\, , \ea \noindent being $\cos\delta =
\sin\alpha^L/\gamma(1-\beta \cos\alpha^L)$ and $\tilde{T}\equiv N_0/\Omega = T \cdot
w/(w^2+(\gamma \beta \cos\delta \,c T)^2)^{1\over 2}$. Expressing
the Dirac  function in Eq. \ref{Yield4} in terms of the
relativistic $\gamma$ factor of the scattering electron(s), we
obtain a close expression {\it which links the scattered photon
yield spectrum to the energy distribution function of the electron
bunch }$F(\gamma)$: \ba \label{Yield5} {d^2 N\over d\omega d
\cos\theta}(\omega,\theta) &=&\alpha{\sqrt{2\pi} \over
8} a^2_0 \int d\theta_e \Theta_{\tilde\gamma}(\theta_e) N_0({\tilde \gamma}) F(\tilde{\gamma}) \times
\nonumber \\ &\times & {\tilde{\gamma}{\cal
H}(\tilde{\gamma}) e^{-{\cal K}(\tilde{\gamma})(\Delta
T-\zeta/(\beta(\tilde{\gamma}c)))^2}\over \Omega
(\cos(\theta-\theta_e)-\cos\alpha^L)} \, , \ea \noindent being
\be
\label{gamma}
\tilde{\gamma} \equiv {(\omega \cos(\theta-\theta_e) - \Omega \cos\alpha^L) \over \sqrt{(\omega \cos(\theta-\theta_e) - \Omega \cos\alpha^L)^2-(\omega-\Omega)^2}}\ee
\noindent the relativistic factor of the electron(s) which generates the scattered photon(s).

We are now able to build up the formula to be used by the energy
spectrometer. Let $Eff(E)$ be the experimental revelation efficiency
of a photon of energy $E$ and $$S(E,\theta)\equiv {dN\over d E
d\cos\theta}|_{Exper}$$ \noindent the detected spectrum of the
photon yield, being $E$ and $\theta$ the  energy and scattering
angle. The energy spectrum of the electron bunch $F(\gamma)$ can
be retrieved by $S(E,\theta)$ and by the prior knowledge of the
size and divergence of the electron bunch as \ba \label{Spectrum1}
F(\gamma) &=& {8 \over \sqrt{2\pi} \alpha
a_0^2N_0(\gamma)}E^{Laser} (1-\cos\alpha^L)\nonumber\\ &\times& {
e^{{\cal K}(\gamma)(\Delta T -\zeta/(\beta(\gamma)c)^2} \over
\gamma {\cal H}(\gamma) } {\cal D}\left({S \over Eff}, \Theta,
\gamma\right) \ea \noindent where ${\cal D}\left({S \over Eff},
\Theta, \gamma\right) $ means the {\it deconvolution } of
$S(E,\theta)/Eff(E)$ with the known angular distribution function
$\Theta_\gamma(\theta_e)$, $ E =
E^{Laser}(1-\beta(\gamma)\cos\alpha)/(1-\beta(\gamma)\cos(\theta-\theta_e))$
and $E^{Laser}=\hbar \Omega$ is the energy of the incoming
photons.

Formulae (\ref{HK}-\ref{Spectrum1}) strongly simplify in the case of very large waist size and    probe pulse incidence angle $\alpha^L$ approaching  $\pi$. In the limit case $w/\sigma_T \rightarrow\infty$, $\alpha^L\rightarrow\pi$, the laser pulse contains all the electron bunch provided that the length of the bunch $\sigma_L$ does not exceeds the probe Reyleigh length $Z_R \approx \pi w^2/\lambda$, so the envelope functions $\cal{H,K}$ reduce to
\be
\label{Limit}
{\cal H}  \rightarrow 1,  \,\,
{\cal K}  \rightarrow 0.
\ee
\noindent In this case the energy spectrum of the electron bunch can be very accurately estimated as
\ba
\label{Spectrum2}
F(\gamma) &=& 2.4\cdot 10^{-2} {w^2\, E^{Laser}\over  {\cal E}\lambda}{\cal D}\left({S \over Eff}, \Theta, \gamma\right)\nonumber \\
E &=& 2E^{Laser}{1\over(1-\beta(\gamma)\cos(\theta-\theta_e))}\, ,
\ea
\noindent where ${\cal E}$ is the energy delivered by the probe pulse (in Joules), $w$ and  $\lambda$ are in $\mu m$, $E^{Laser}$ is in $eV$  and we have assumed $\theta <<1$ and $\gamma>>1$.

In order to present a full simulation of a possible experimental setup, we will focus on the measure of the spectrum of a relativistic electron bunch produced e.g.  by Laser-Plasma acceleration (say Laser Wake Field Acceleration, LWFA, or Self-Modulated-LWFA) \cite{LWFA}\cite{Dawson}.  In this framework, an ultraintense laser pulse inpinges onto a plasma and the strong electric fields of the wake accelerate a large number of electrons at energies exceeding tens of MeV's \cite{Nostro}. Usually the energy spectra of the electron bunches are broad and, in the cases of acceleration with uncontrolled trapping, an exponentially decreasing distribution is obtained \cite{Leemans}. Attempts to control the trapping of the accelerated electrons, thus producing electron buches with low spectral dispersion, are in progress \cite{Leemans1} \cite{Controlled1}.

 To face  with a realistic electron bunch, we  simulated  $N_e = 6\,
 10^8$ electrons with Gaussian angular distribution
 $\Theta_\gamma(\theta_e)$ having divergence ranging from $20 mrad$ at
 low energies to $10 mrad$ at large energies. The spatial distribution has transverse and longitudinal size $\sigma_T = 20 \mu m$, $\sigma_L = 50 \mu m $, respectively.
Finally, the  energy distribution $F(\gamma)$ is composed  both by
an exponentially decreasing background (whose parameters have been
estimated by experimental data \cite{Nostro}) and a few MeV's
thick Gaussian peak, taking into account the portion of the bunch
(possibly) generated with controlled trapping. The probe pulse
propagates against the simulated electron bunch ($\alpha^L =
\pi$), is $ 1 ps$ long, has energy ${\cal E} = 0.1 J$  and waist
size $w = 50 \mu m$.

\begin{figure}
\includegraphics[width=5in]{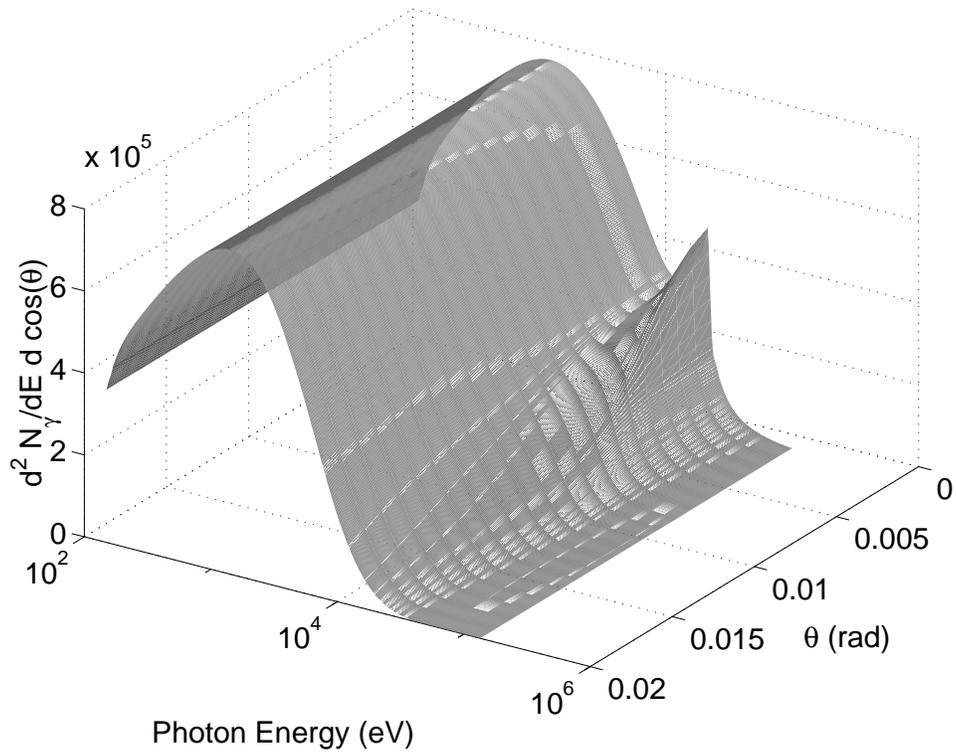}
\caption{Numerical simulation of the scattered photon yield integrated onto the azimuthal angle $\phi$. \label{fig:yield}}
\end{figure}

The simulation of the detected photon yield is performed with a Monte Carlo method by using Eq.  \ref{Yield1}, assuming an acceptance angle of the photon detector (a CCD camera, see below) of $30 mrad$ and a detection efficiency  in the range $0.2-1$ (see \cite{CCDeff}). The angular and spectral distribution of the $\approx 10^6 $ detected  photons  is shown in Fig. \ref{fig:yield}.

\begin{figure}
\includegraphics[width=5in]{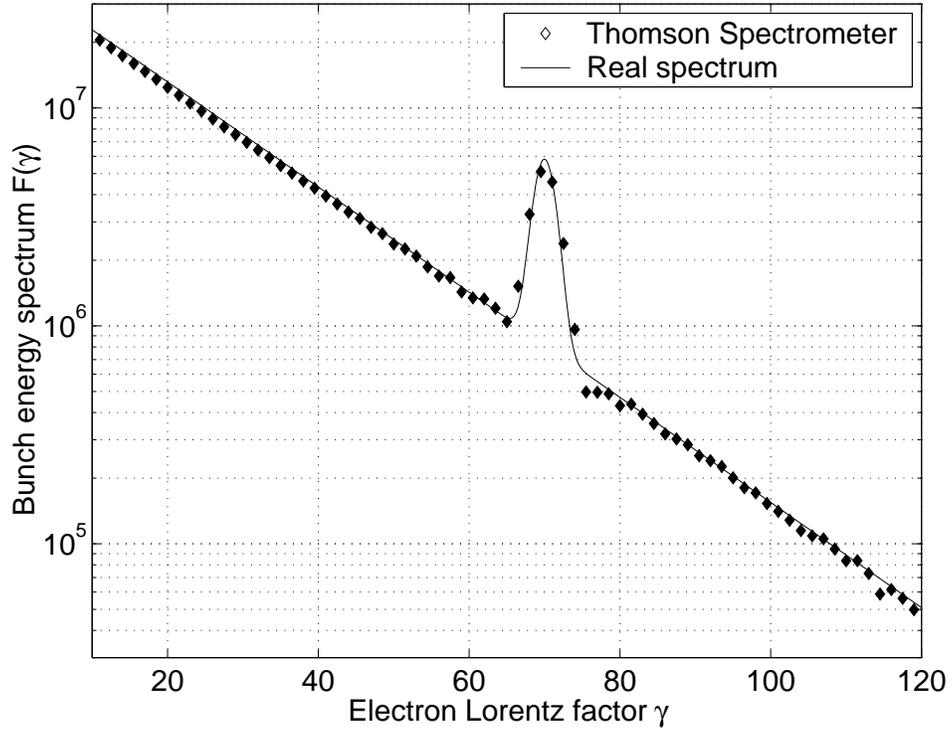}
\caption{Comparison between the bunch energy spectrum estimated with the Thomson spectrometer and the simulated spectrum (full line). \label{fig:Results}}
\end{figure}

Since the probe and bunch parameters allow the application of the simplified analysis (see Eq. \ref{Limit}),  to analyze the 'experimental' results by using Eq. \ref{Spectrum2} we  proceed by computing the deconvolution of the photon yield with the angular distribution of the bunch. The result of the analysis  is shown in Fig. \ref{fig:Results}.  It is clear that the estimated energy spectrum well reproduces  the simulated one also at large energies, where the effect of the bunch divergence is significant. We note also that the results of the analysis of the sumulated data clearly show that the electron spectrometer is capable of measuring, in a single shot, the  very wide energy spectrum with a resolution good enough to detect a narrow peak.

\begin{figure}
\includegraphics[width=2.5in]{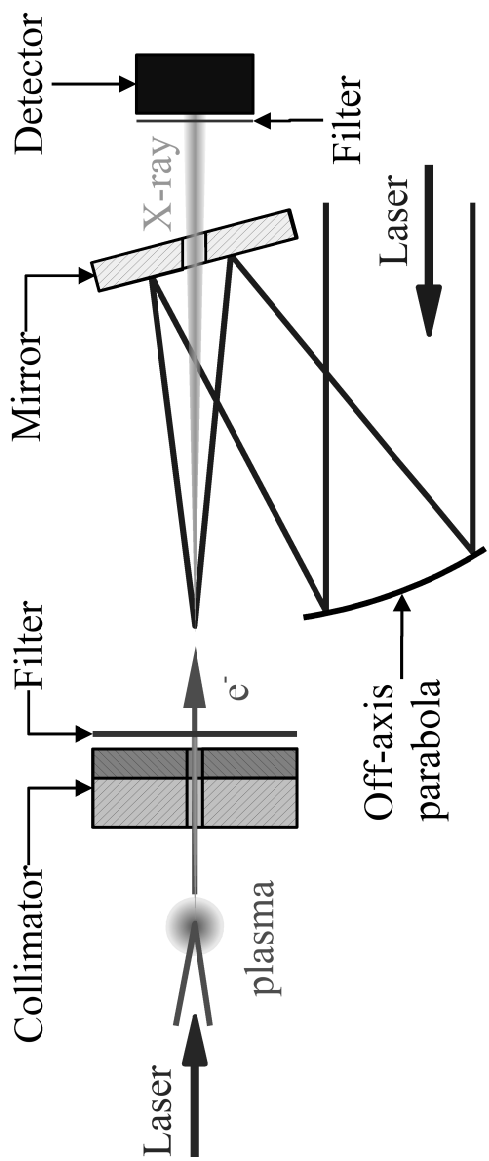}
\caption{Example of a possible experimental setup. The probe pulse
is focussed onto the electron bunch with an off-axis parabola and
the scattered photons are detected with an ADC-calibrated CCD
camera.\label{fig:setup2}}
\end{figure}

 A possible setup of the spectrometer is shown in Fig. \ref{fig:setup2}.
The probe pulse is focused with an off-axis parabola and directed with a mirror towards the electron bunch. The Thomson scattered radiation is detected with an energy  calibrated CCD camera working in a single-photon mode \cite{SinglePhoton}.
 In order to ensure the acquisition of a large number of photons for each shot, the CCD camera should have both a large number of pixel and a reasonable high quantum efficiency ($QE$) at large energies \cite{CCDeff}.

Finally, we briefly discuss the main sources of uncertainty of the
electron spectrometer. Let $(\gamma_{min}, \gamma_{Max})$ be
the range of measure of the electron spectrometer, and $\Delta E$, $\delta(\Delta\theta_e)$ the uncertainties on the
energy of a scattered photon and on the bunch divergence, respectively. Taking into account Eq. \ref{gamma},
we get
\be
\label{Error1}
{\delta \gamma\over \gamma}\approx {1\over 2}
{\Delta E\over E}\approx {1\over 2}\left({1\over \gamma^2}{\Delta E_{Exp}\over 4 E^{Laser}}+{1\over \Omega {\tilde T}} +\gamma^2{ \delta(\Delta\theta_e)^2\over 4}\right)
\ee
\noindent provided that $\delta(\Delta\theta_e) << 1/\gamma_{Max}$. Eq. (\ref{Error1}) shows that the range of measurable energy spectrum {\it is limited on the low energy side by the energy resolution of the photon detector and on the high energy side  by the error on the beam divergence}. Supposing reasonable values for
$\Delta E_{Exp}$ and $\delta(\Delta\theta_e)$ of $100 eV$ and $0.5 mrad$ (see Ref. \cite{Leemans1}), we obtain that estimation of the $\gamma$ factor are avaible within (say) $5$ percent error in the very wide band $10-1000$.
The uncertainty on the amount of electrons having $\gamma$ fixed  depends critically on the experimental setup. Apart from counting errors, strong error sources are the uncertainty on the spatial  bunch parameters, while errors in the estimations of the laser parameters are usually low. The main uncertainty can be, in fact,  on the envelope of the intersection between the electron and laser beams, which is described by the functions ${\cal H, \, K}$. We stress immediately that, if the experimental environment let the laser pulse counterpropagate against the electron bunch ($\alpha^L = \pi$), this  error source can be made negligible, provided that the laser waist is large enough (see Eqq. (\ref{HK},\ref{Limit})). In this case the largest error sources on $F(\gamma)$ should realistically be linked to the statistics on the photon counting, so if a {\it single shot} electron spectrometer is wanted, a large number of photons for each shot should be detected, as in the example just analyzed.

To conclude, we have shown that Thomson Scattering of a laser beam onto a relativistic electron bunch at non relativistic laser intensities can be used to generate a beam of scattered photons and that the distribution of the  photon yield can be linked to the energy distribution of the electron bunch. In the case of probe waist size much higher than the transversal size of the counterpropagating electron beam, a strong simplification of the analysis of the experimental data occurs and a very accurate spectrometer can been implemented.

One of the
 authors (PT) wish to acknowledge support from  Italian M.I.U.R (Project ``Metodologie e diagnostiche per materiali e ambiente).

\newpage

\newpage

\newpage

\begin{thebibliography}{99}
\bibitem{Leemans1}
{W.P. Leemans {\it et. al.}, PRL {\bf 77} 20, 4182-4185 (1996) }
\bibitem{Leemans2}
{W.P. Leemans {\it et. al.}, PRL {\bf 67} 11, 1434-1437 (1991) }
\bibitem{Renard}
{N. Renard {\it et. al.}, PRL {\bf 77} 18, 3807-3810 (1996)}
\bibitem{LaserPlasma}
{K. Krushelnick {\it et. al.}, PRL {\bf 78} 21, 4047-4050 (1997)}
\bibitem{fus1}
{T.N. Carlstrom {\it et. al.}, Rev. Sci. Instrum. {\bf 63} 10, 4901-496 (1992)}
\bibitem{fus2}
{K. Muraoka, K. Uchino and M.D. Bowden, Plasma Phys. Control. Fusion {\bf 40}, 1221-1239 (1998)}
\bibitem{Leemans}
{P. Catravas, E. Esarey and W. P. Leemans, Meas. Sci. Technol. {\bf 12} (2001) 1828-1834}
\bibitem{Medical}
{A. Ts. Amatuni and M. L. Petrossian, Nucl. Meth. in Phys. Res. {\bf A 455}, 128-129 (2000)}
\bibitem{Sp1}
{J. J. Livingood,, {\it The Optics of Dipole Magnets}, Academic press N.Y. and London, 1969 }
\bibitem{Nostro}
{D. Giulietti {\it et. al.}, submitted to Phys. of Plasmas, 2002}
\bibitem{LWFA}
{D. Gordon {\it et. al.}, PRL {\bf 80}, 2133 (1998) }
\bibitem{Jackson}
{J.D. Jackson, {\it Classical Electrodynamics}, (Wiley, New York, 1975}
\bibitem{Esarey}
{E. Esarey, S. K. Ride and P. Sprangle, PRE {\bf 48}, 4 3003-3021
 (1993) }
\bibitem{Ride}
{S. K. Ride, E. Esarey and M. Baine, PRE {\bf 52}, 5 5425-5442 (1995) }
\bibitem{Dawson}
{T. Tajima and J. M. Dawson, PRL {\bf 43}, 267 (1979) }
\bibitem{Controlled1}
{ S. Bulanov, N. Naumova, F. Pegoraro and J. Sakai, PRE {\bf 58} 5, R5257-R5260 (1998)}
\bibitem{SinglePhoton}
{L.A. Gizzi {\it et. al.}, PRL {\bf 76} 2278 (1996) }
\bibitem{CCDeff}
{C.M. Castelli and G.W. Fraser, Nucl. Instr. Meth. in Phys. Res. A {\bf 376} (1996) 298-300}

\end{thebibliography}
\end{document}